\documentclass[a4paper,11pt]{article}
\pdfoutput=1

\usepackage{array}
\usepackage{amsfonts}
\usepackage{amsmath}
\usepackage{amssymb}
\usepackage[small]{caption}
\usepackage{cite}
\usepackage[top=1.2in, bottom=1.2in, left=1.1in, right=1.1in]{geometry}
\usepackage{graphicx}
\graphicspath{{./}{./figures/}}
\usepackage{pifont}
\newcommand{\xmark}{\ding{53}}%
\usepackage{slashed}
\usepackage{color}

\definecolor{mediumblue}{rgb}{0,0,0.8}
\usepackage{hyperref}
\hypersetup{
  linktocpage=true,
  colorlinks=true,
  citecolor=mediumblue,
  filecolor=mediumblue,
  linkcolor=mediumblue,
  urlcolor=mediumblue,
}
\newcommand{\mailref}[1]{\href{mailto:#1}{#1}}

\allowdisplaybreaks

\def\l{\ensuremath{\left}}
\def\r{\ensuremath{\right}}

\newcommand{\I}{\ensuremath{i\mkern1mu}}

\newcommand{\mc}{\mathcal}
\newcommand{\nno}{\nonumber}
\newcommand{\fr}{\frac}

\begin{document}

\begin{titlepage}
\def\thefootnote{\fnsymbol{footnote}}

\begin{flushright}
  \texttt{CTPU-PTC-18-28}\\
  \texttt{EFI-18-13}
\end{flushright}

\begin{centering}
\vspace{0.5cm}
{\Large \bf \boldmath
  Collider probes of singlet fermionic dark matter scenarios\\[0.2cm]
  for the Fermi gamma-ray excess}

\bigskip

\begin{center}
  {\normalsize 
    Yeong~Gyun~Kim$^{a,}$\footnote{\mailref{ygkim@gnue.ac.kr}},
    Chan~Beom~Park$^{b,}$\footnote{\mailref{cbpark@ibs.re.kr}}, and
    Seodong~Shin$^{c,d,}$\footnote{\mailref{seodongshin@yonsei.ac.kr}}}\\[0.5cm]
  \small
  $^a${\em Department of Science Education, Gwangju National
    University of Education,\\
    Gwangju 61204, Korea}\\[0.1cm]
  $^b${\em Center for Theoretical Physics of the Universe,\\
    Institute for Basic Science (IBS), Daejeon 34051, Korea}\\[0.1cm]
  $^c${\em Enrico Fermi Institute, University of Chicago, Chicago, IL
    60637, USA}\\[0.1cm]
  $^d${\em Department of Physics and IPAP, Yonsei University, Seoul
    03722, Korea}
\end{center}
\end{centering}

\medskip

\begin{abstract}
  \noindent
  We investigate the collider signatures of the three benchmark points in the
  singlet fermionic dark matter model.
  The benchmark points, which were introduced previously to explain
  the Fermi gamma-ray excess by dark matter (DM) pair annihilation at
  the Galactic center, have definite predictions for future collider
  experiments such as the International Linear Collider and the
  High-Luminosity LHC\@.
  We consider four collider observables: (1) Higgs signal strength
  (essentially $hZZ$ coupling), (2) triple Higgs coupling, (3) exotic
  Higgs decay, and (4) direct production of a new scalar particle.
  The benchmark points are classified by the final states of the DM
  annihilation process: a pair of $b$ quarks, SM-like Higgs bosons,
  and new scalar particles.
  Each benchmark scenario has detectable new physics signals for the
  above collider observables that can be well tested in the future
  lepton and hadron colliders.
\end{abstract}

\vspace{0.5cm}

\end{titlepage}

\renewcommand{\thefootnote}{\arabic{footnote}}
\setcounter{footnote}{0}

\setcounter{tocdepth}{2}
\noindent \rule{\textwidth}{0.3pt}\vspace{-0.4cm}\tableofcontents
\noindent \rule{\textwidth}{0.3pt}

\section{Introduction\label{sec:intro}}

The existence of dark matter (DM) in the universe is now well
established by various experiments observing the gravitational
interactions of the DM and the anisotropy of cosmic microwave background.
In order to investigate its feature as a particle, it is required to
observe non-gravitational interactions of DM with the standard model
(SM) particles.
One of such trials is indirect detections of the DM, which probe the
signals from the annihilation or the decay of the DM in the current universe.
They can contribute to energetic charged particles, photons, and
neutrinos which are observable in satellite and
terrestrial detectors.
In particular, the gamma-ray signals have always drawn attention in
the sense that we can identify the location of the sources.
Interestingly, several independent researches have reported an excess
of the gamma-ray emission from the Galactic center (GC) above the
expected astrophysical background from the analysis of the Fermi Large
Area Telescope (LAT)
data~\cite{Goodenough,Hooper:2010mq,Hooper:2011ti,Abazajian:2012pn,
  Hooper:2013rwa,Gordon:2013vta,Huang:2013pda,Abazajian:2014fta,
  Daylan:2014rsa,Calore:2014xka,Calore:2014nla}, confirmed later by the experimental group \cite{TheFermi-LAT:2015kwa}.

The excess can be explained by DM annihilations or decays although the
explanations by unidentified astrophysical sources still remain viable
possibilities.
In Ref.~\cite{Kim:2016csm} we examined the possibility that the GeV
scale Fermi gamma-ray excess at the GC can be explained by the DM
annihilation in the singlet fermionic dark matter (SFDM)
model~\cite{Kim:2008pp, Kim:2009ke}.
Within the framework, we showed that the DM annihilation into
a bottom-quark pair, Higgs pair, and new scalar pair can provide good
fits to the Fermi gamma-ray excess data.
However, due to the unknown populations of other astrophysical sources
near the GC, it is very hard to confirm that the excess is originated
from the DM only via the astrophysical observations in general.
Hence, as a complementary approach to support the DM explanation, it
is necessary to probe the relevant DM scenarios in collider
experiments where the background events are relatively well
controlled.
Following this strategy, we investigate the detection prospects of the
SFDM signals explaining the gamma-ray excess in future colliders such
as the High-Luminosity Large Hadron Collider (HL-LHC) and the
International Linear Collider (ILC) as reference machines of hadron
and lepton colliders.
Note that the analysis results in the ILC can be easily converted to
those in the Circular Electron Positron Collider (CEPC) as well.

The reference model parameters in Ref.~\cite{Kim:2016csm} are chosen
to explain the Fermi gamma-ray excess with a best fit for each
annihilation channel: resonant $b \bar b$ production, double Higgs
production, Higgs and a new scalar production when they are almost
degenerate in mass, and double new scalar production.
Then, we can proceed analyses with these fixed reference parameters
providing definite predictions for collider phenomenology.
The scenario, where the SM-like Higgs boson and the new scalar are too
degenerate in mass and the couplings of the new scalar to $WW$ and
$ZZ$ are highly suppressed, is quite hard to be probed in collider
experiments.
Thereby we only consider the reference parameters of the other three channels
and name them benchmark point (BP) I, II, and III, in order.

This paper is organized as follows. We briefly describe the SFDM model
in Sec.~\ref{sec:model}.
The benchmark scenarios for explaining the Fermi gamma-ray excess are
introduced and their possible collider signatures in the future
collider experiments are discussed in Sec.~\ref{sec:bench}.
Section~\ref{sec:concl} is devoted to conclusions.

\section{Singlet fermionic dark matter}
\label{sec:model}

In this section, we summarize the key features of the SFDM model used
in explaining the gamma-ray excess from the GC~\cite{Kim:2016csm}.
The dark sector is composed of a real scalar field $S$ and a Dirac
fermion field $\psi$, both of which are singlet under the SM gauge group.

The Lagrangian for the dark sector is given by the following
renormalizable interactions.
\begin{align}
  \mc{L}^{\rm dark} = \bar\psi (\I \slashed{\partial}
   - m_{\psi_0}) \psi + \fr{1}{2} \partial_\mu S \partial^\mu S  - g_S
  (\cos\theta \,\bar\psi \psi + \sin\theta \,\bar\psi \I \gamma^5 \psi) S
  - V_S (S, \, H),
  \label{eq:lagrangian}
\end{align}
where the singlet scalar potential is
\begin{align}
  V_S (S, \, H) = \fr{1}{2} m_0^2 S^2 + \lambda_1 H^\dagger H S +
  \lambda_2 H^\dagger H S^2 + \fr{\lambda_3}{3!} S^3 +
  \fr{\lambda_4}{4!} S^4 .
\end{align}
As compared to the original proposal of the SFDM~\cite{Kim:2008pp},
the pseudoscalar interaction in the dark sector is further introduced
to obtain a good fit to the gamma-ray excess from the GC using the DM
annihilation.
See Ref.~\cite{Kim:2016csm} for the detail of the fit to the gamma-ray
data.

The SM Higgs potential is given as
\begin{align}
  V_\mathrm{SM} = -\mu^2 H^\dagger H + \lambda_0 (H^\dagger H)^2 .
\end{align}
The Higgs doublet $H$ is written in the unitary gauge after the
electroweak symmetry breaking as follows:
\begin{equation}
  H = \fr{1}{\sqrt{2}}
  \begin{pmatrix}
    0 \\ v_h + h
  \end{pmatrix}
\end{equation}
with $v_h \simeq 246$~GeV.
The singlet scalar field also develops a nonzero vacuum expectation
value, $v_s$, and the singlet scalar field is written as $S = v_s +
s$.
The mass parameters $\mu^2$ and $m_0^2$ can be expressed in terms of
other parameters by using the minimization condition of the full
scalar potential $V_S + V_\mathrm{SM}$, \textit{i.e.},
\begin{align}
  \mu^2 &= \lambda_0 v_h^2 + (\lambda_1 + \lambda_2 v_s) v_s ,\nno\\
  m_0^2 &= - \l( \fr{\lambda_1}{2 v_s} + \lambda_2 \r) v_h^2
          - \l( \fr{\lambda_3}{2 v_s} + \fr{\lambda_4}{6} \r) v_s^2 .
\end{align}
The mass terms of the scalar fields are
\begin{equation}
  -\mc{L}_\mathrm{mass} =
  \fr{1}{2} \mu_h^2 h^2 + \fr{1}{2} \mu_s^2 s^2 + \mu_{hs}^2 h s ,
  \label{eq:higgs_mass_matrix}
\end{equation}
where
\begin{align}
  \mu_{h}^2 &= 2\lambda_0 v_h^2 ,\nno\\
  \mu_{s}^2 &= -\fr{\lambda_1 v_h^2}{2 v_s} + \fr{(3\lambda_3 +
              2\lambda_4 v_s) v_s}{6} ,\nno\\
  \mu_{hs}^2 &= (\lambda_1 + 2\lambda_2 v_s) v_h .
  \label{eq:baremass}
\end{align}
A non-vanishing value of $\mu_{hs}^2$ induces mixing between the SM Higgs field configuration $h$ and the singlet scalar field $s$ as
\begin{equation}
  \begin{pmatrix}
    h_1 \\ h_2
  \end{pmatrix} = \begin{pmatrix}
    \cos\theta_s & \sin\theta_s\\
    -\sin\theta_s & \cos\theta_s
  \end{pmatrix} \begin{pmatrix}
    h \\ s
  \end{pmatrix} ,
\end{equation}
The mixing angle $\theta_s$ is given by
\begin{equation}
  \tan\theta_s = \fr{y}{1 + \sqrt{1 + y^2}}
\end{equation}
with $y \equiv 2\mu_{hs}^2 / (\mu_h^2 - \mu_s^2)$.
Then, the physical Higgs boson masses are
\begin{equation}
  m_{h_1,\,h_2}^2 = \fr{1}{2} \l[ (\mu_h^2 + \mu_s^2) \pm (\mu_h^2 -
  \mu_s^2) \sqrt{1 + y^2} \r].
\end{equation}
We define that $h_1$ is the SM-like Higgs boson with $m_{h_1} =
125$~GeV and $h_2$ is the singlet-like scalar boson throughout this
paper.

The imaginary mass term of the DM particle $\psi$ from the
pseudoscalar interaction, proportional to $\sin\theta$ in
(\ref{eq:lagrangian}), can be eliminated by a chiral transformation
and field redefinition, as
stated in Refs.~\cite{Kim:2016csm,Fedderke:2014wda}.
Then, one can find that the DM mass is given by
\begin{equation}
  m_\psi  = \sqrt{(m_{\psi_0} + g_S v_s\cos\theta)^2 + g_S^2 v_s^2
    \sin^2\theta} .
\end{equation}
And the dark sector Yukawa interactions are redefined as
\begin{align}
  -\mc{L}^{\rm dark}_\mathrm{int} = g_S \cos\xi \,s \bar\psi \psi + g_S \sin\xi \,s
  \bar\psi \I \gamma^5 \psi ,
\end{align}
where
\begin{align}
  \cos\xi &= \fr{m_{\psi_0} \cos\theta + g_S v_s}{m_\psi} ,\nno\\
  \sin\xi &= \fr{m_{\psi_0} \sin\theta}{m_\psi} .
\end{align}
Therefore, there are three independent model parameters for
the singlet fermion: $m_\psi$, $g_S$, and $\xi$.
On the other hand, the masses $m_{h_1, \, h_2}$, the mixing angle
$\theta_s$, and self-couplings of the two physical Higgs particles
$h_1$ and $h_2$ are determined by the six independent parameters in
the scalar potential, $\lambda_0$, $\lambda_1$, $\lambda_2$,
$\lambda_3$, $\lambda_4$, and $v_s$.
For a reference, we recall that the scalar triple Higgs self-couplings
$c_{ijk}$ for $h_i h_j h_k$ interactions are expressed as
\begin{align}
  c_{111}
  =&~ 6 \lambda_0 v_h \cos^3 \theta_s + \l( 3 \lambda_1 + 6 \lambda_2
     v_s \r) \cos^2 \theta_s \sin\theta_s + 6 \lambda_2 v_h
     \cos\theta_s \sin^2\theta_s + (\lambda_3 + \lambda_4 v_s)
     \sin^3\theta_s , \nno\\
  c_{112}
  =& -6 \lambda_0 v_h \cos^2 \theta_s \sin\theta_s + 2 \lambda_2 v_h
     \l( 2 \cos^2\theta_s \sin\theta_s - \sin^3 \theta_s\r) \nno\\
  &  + \l(\lambda_1 + 2 \lambda_2 v_s \r) \l( \cos^3 \theta_s -
    2 \cos\theta_s \sin^2\theta_s \r) + \l( \lambda_3 + \lambda_4 v_s
    \r) \cos\theta_s \sin^2\theta_s , \nno\\
  c_{122}
  =&~ 6 \lambda_0 v_h \cos \theta_s \sin^2 \theta_s + 2 \lambda_2 v_h
     \l( \cos^3\theta_s - 2 \cos\theta_s \sin^2 \theta_s\r) \nno\\
  &  - \l(\lambda_1 + 2 \lambda_2 v_s \r) \l( 2 \cos^2 \theta_s
    \sin\theta_s - \sin^3 \theta_s \r) + \l( \lambda_3 +
    \lambda_4 v_s \r) \cos^2\theta_s \sin\theta_s , \nno\\
  c_{222}
  =& - 6 \lambda_0 v_h \sin^3 \theta_s + \l( 3 \lambda_1 + 6
     \lambda_2 v_s \r) \sin^2 \theta_s \cos\theta_s - 6 \lambda_2 v_h
     \sin\theta_s \cos^2\theta_s \nno\\
  & + (\lambda_3 + \lambda_4 v_s) \cos^3\theta_s .
\label{eq:cijk}
\end{align}

\section{Benchmark points and collider signatures\label{sec:bench}}

We consider the three benchmark points of the SFDM model, which can
explain the Fermi gamma-ray excess from the DM pair annihilations.
In Ref.~\cite{Kim:2016csm}, we demonstrated in detail that the
annihilation of DM into a bottom-quark pair, Higgs pair, and new
scalar pair can give good fits to the Fermi-LAT gamma-ray data.
Because model parameters are fixed in order to explain the Fermi
gamma-ray excess, it gives rise to definite predictions for collider
phenomenology.
If the predictions are confirmed by the future collider experiments, the benchmark points will be strongly favored as the
solution for the Fermi gamma-ray excess.

The benchmark points we consider for the collider study are classified
by the final state of the main DM annihilation process.
The first benchmark point corresponds to the DM annihilation channel
into a bottom-quark pair (BP I), while the second and third benchmark points correspond to the DM
annihilation channels into a Higgs pair (BP II) and into a new scalar pair (BP III), respectively, which subsequently decay to the SM particles.
Let us now discuss the details of the three benchmark points and their
collider signatures one by one.

\subsection{BP I (\boldmath$\psi \bar{\psi} \rightarrow b \bar{b}$ annihilation channel)}

The channel of DM annihilation into a bottom-quark pair is one of the
most widely considered possibilities explaining the gamma-ray excess.
For example, the model independent study in Ref.~\cite{Calore:2014nla}
showed that the DM annihilation into $b\bar{b}$ gives a good fit to
the gamma-ray excess data if $m_\text{DM} \simeq 48.7$~GeV and the DM
annihilation cross section $\langle \sigma v \rangle \simeq 1.75\times
10^{-26}$~$\mathrm{cm}^3\,\mathrm{s}^{-1}$ for a self-conjugate DM\@.
In the SFDM model, this channel can give a best fit to the observed data
when $m_\psi = 49.706$~GeV, $m_{h_2} = 99.416$~GeV, $m_{h_1} =
125.3$~GeV, and $\sin\theta_s = - 0.117$ from the reference parameters
$\lambda_0 = 0.128816$, $\lambda_1 =36.625338$~GeV, $\lambda_2 = -
0.131185$, $\lambda_3 = - 333.447606$~GeV, $\lambda_4 = 5.648618$,
$v_s = 150.017297$~GeV, $g_S = 0.055$, and $\sin\xi
=0.01$~\cite{Kim:2016csm}.
With the parameter setup, we obtain the correct DM relic density
$\Omega h^2 = 0.118$, and the DM annihilation cross section $\langle
\sigma v \rangle = 1.5 \times
10^{-26}$~$\mathrm{cm}^3\,\mathrm{s}^{-1}$ which can explain the Fermi
gamma-ray excess within the uncertainty of the galaxy halo profile
near the GC\@.
Note, however, that only a narrow parameter region around the
resonance of $\psi \bar \psi \to b \bar b$ is allowed to fit the correct
relic density and the new scalar $h_2$ needs to be almost scalar in
the dark sector, {\em i.e.}, $\sin\xi = 0.01$, to avoid the strong
astrophysical bounds from the observation of the gamma-rays coming
from the dwarf spheroidal galaxies by
Fermi-LAT~\cite{Ackermann:2015zua} and the antiproton ratio by
PAMELA~\cite{Adriani:2010rc} and AMS-02~\cite{Aguilar:2016kjl}.
On the other hand, the small mixing angle between the SM Higgs and the
singlet scalar, {\em i.e.}, $\sin\theta_s = - 0.117$ suppresses the
spin-independent cross section of the DM recoiling against nucleon as
$\sim 6.3 \times 10^{-48}~\mathrm{cm}^2$, which is below the
constraints from various DM direct detection experiments.
In the following subsections, we discuss the collider signatures of
this benchmark point in terms of the Higgs signal strength, triple
Higgs coupling, exotic Higgs decays, and direct production of $h_2$.

\subsubsection{Higgs signal strength\label{subsec:higgs_signal_strengh}}

As already mentioned, the physical Higgs states are admixtures of $h$
and $s$ in the SFDM model.
Therefore, the SM-like Higgs couplings to SM gauge bosons and
fermions are universally suppressed by the factor of $\cos\theta_s$,
compared to the couplings in the SM\@.

The universal reduction factor $\cos\theta_s$ can be precisely measured
at the ILC\@.
The SM cross section for the Higgsstrahlung process $e^+ e^-
\rightarrow Zh$ reaches its maximum value at $\sqrt{s}=250$~GeV.
About a half million $Zh$ events are expected from the integrated
luminosity of 2~ab$^{-1}$ with a suitable polarization of $e^+ e^-$
beams.
Then, it is possible to precisely measure the inclusive cross
section of the Higgsstrahlung process using the recoil mass technique,
which disregards the decay products of the Higgs boson.
For a $Z$ boson decaying into a pair of fermions, $Z \to f \bar f$,
the recoil mass is defined as~\cite{Fujii:2015jha, Liu:2017lpo}
\begin{equation}
  M_\text{recoil} = \sqrt{s - 2 \sqrt{s} (E_f + E_{\bar f}) + m_{f
      \bar f}^2} .
\end{equation}
It corresponds to the mass of the Higgs boson associated with the $Z$
boson in the Higgsstrahlung process.
On the other hand, the simplest approach to represent the effect of
new physics is the so-called $\kappa$ formalism, where the form of
Higgs interactions to the SM particles are the same as the SM, but the
couplings are rescaled from the SM value.
In the SFDM model, the cross section $\sigma_{Zh_{1}}$ is proportional
to $\cos^2\theta_s$.
The $\cos\theta_s$ value can be determined very precisely, with
accuracy of 0.38\% assuming a total integrated luminosity of
2~ab$^{-1}$ in the $\kappa$ formalism:
\begin{equation}
  \kappa_Z^2 = \frac{\sigma (e^+ e^- \to Z h_1)}{\sigma (e^+
    e^- \to Z h)} = \cos^2\theta_s ,
\end{equation}
where the denominator is the SM prediction.
Higher energy stages of the ILC experiments will reach an accuracy of
0.3\% for the $hZZ$ coupling~\cite{Fujii:2017vwa}.\footnote{
  Furthermore, it is expected that the FCC-ee experiments at 240~GeV
  (5~ab$^{-1}$) and at 365~GeV (1.5 ab$^{-1}$), combined with
  measurements of single and double Higgs processes at the HL-LHC,
  can achieve $\sim 0.25$\% of the accuracy for $hZZ$
  coupling~\cite{DiVita:2017vrr}.}


The mixing angle $\sin\theta_s = -0.117$ of the BP I
implies 0.7\% deviation of the $hZZ$ coupling from the SM value.
Therefore the ILC experiment will be able to measure the
deviation. On the other hand, the HL-LHC experiment is expected to
have an accuracy of $\sim 4$\%~\cite{CMS:2013xfa, ATLAS:HL-LHC-Higgs},
which is insensitive to the deviation of the Higgs coupling.

\subsubsection{Triple Higgs coupling}

Another possible deviation from the SM couplings comes from the triple
Higgs self-coupling $c_{111}$.
In the SFDM model, the Higgs self-couplings are given by Eq.~(\ref{eq:cijk}).
The prospects for measuring the Higgs self-coupling is not so
promising at the LHC, even at the HL-LHC\@.
Only $\mathcal{O}(1)$ accuracy is expected for the triple Higgs
coupling $c_{111}$ from the observation of the Higgs pair production in
the channel $hh \to b \bar b \gamma \gamma$ at the HL-LHC~\cite{ATLAS:2017-001}.
Meanwhile, projections for 100~TeV $pp$ colliders show that a very
good precision on the determination of the triple Higgs coupling, a
statistical precision of the order of 4\%, is achievable using the $b
\bar b \gamma \gamma$ channel with the integrated luminosity of
30~ab$^{-1}$~\cite{He:2015spf,Contino:2017}.
For the BP I, the triple Higgs coupling $c_{111}$ is 183.95~GeV, which
corresponds to 3.9\% reduction from the SM value, so it is on the
sensitivity limit of the 100~TeV $pp$ collider.

As for the ILC, high-energy machines with center-of-mass energies above
350~GeV can provide the opportunity of directly probing the coupling
$c_{111}$ through Higgs-pair production processes, in particular,
the double Higgsstrahlung $e^+e^- \rightarrow Z h h$ and $WW$-fusion
$e^+e^- \rightarrow \nu\bar{\nu}h h$ processes.
It is known that the interference between diagrams with and without a
triple Higgs vertex has opposite sign in double Higgsstrahlung and
$WW$-fusion, so that a combination of double Higgsstrahlung and
$WW$-fusion measurements could be used to maximize the precision for
the deviation of the triple Higgs coupling. ILC runs at 500~GeV or
higher energies maximize the overall precision allowing for a
determination of $c_{111}$ with a $\sim 20$\% uncertainty at 68\% C.L.~\cite{DiVita:2017vrr}

\subsubsection{Exotic Higgs decays}

As mentioned in Subsec.~\ref{subsec:higgs_signal_strengh}, the ILC can measure
the absolute size of the inclusive Higgs production cross section
$\sigma(e^+ e^- \rightarrow Z h_1)$ by applying the recoil mass
technique, which is independent of the Higgs decay modes.
The recoil mass technique is applicable even if the Higgs decays
invisibly and hence indispensable for extracting the Higgs branching
ratio.
For the sensitivity to invisible decay modes of the Higgs boson, the
250~GeV ILC with 2~ab$^{-1}$ luminosity and polarized beams would
provide an upper limit $\text{BR}(h \to \text{invisible}) < 0.3$\% at 95\% C.L.~\cite{Barklow:2017suo},
which is a factor of 20 below the expected sensitivity of the HL-LHC~\cite{CMS:2013xfa, ATLAS:LH-LHC-Higgs-invisible}.

The SFDM model can lead to new Higgs decay channels if kinematically allowed.
In the BP I, the mass of the singlet fermion is $m_\psi \simeq 50$~GeV so that the SM-like Higgs $h_1$ can decay into a pair of DM particles,
which will escape the detector without leaving tracks, thus leading to an invisible Higgs decay.
The branching ratio of the invisible Higgs decay is predicted to be
BR$(h_1\rightarrow\psi\bar{\psi}) = 0.73$\%.
Hence, we expect that the invisible Higgs decay of the BP I is within
the reach of the ILC, while it is beyond the reach of the HL-LHC\@.

\subsubsection{Production of {\boldmath$h_2$} at colliders}

The new particles can be produced at colliders, thus enabling
a direct probe of the SFDM model through dedicated search channels.
The search for the light additional Higgs boson $h_2$ ($m_{h_2} < m_{h_1}$)
directly produced via the gluon-gluon fusion process at hadron
colliders suffers from a huge amount of backgrounds if
$h_2 \to b \bar b$ is the dominant decay mode.
Nevertheless, the LHC increases the sensitivity on the search for the
light Higgs by combining with the other production channels.
For instance, the CMS collaboration studied a low-mass resonance
in the diphoton channel by combining the 8~TeV and 13~TeV
data~\cite{CMS:diphoton}~\footnote{See also Ref.~\cite{Liu:2018xsw} and the references therein.}.
The analysis result sets the upper limit to the ratio of the
production cross sections as
\begin{equation}
  \fr{\sigma (pp \to h_2 \to \gamma\gamma)}{\sigma (pp \to h \to \gamma
    \gamma)_\text{SM}} = \sin^2 \theta_s \lesssim 0.25
\end{equation}
at 95\% C.L. for $m_{h_2} = m_h \simeq 99$~GeV.
For the HL-LHC, we expect that the upper limit will be improved by
order of magnitude, $\sin^2 \theta_s \lesssim \mc{O}(0.01)$, so that
the production of $h_2$ can be marginally discovered.

The strongest bound on $\sin^2 \theta_s$ from the direct production of
$h_2$ still comes from LEP\@.
The LEP experiments provide the 95\% C.L. upper bound on
the Higgs mixing angle $\sin^2\theta_s$ as a function of the light
Higgs mass $m_{h_2}$, which corresponds to $\sin^2\theta_s\simeq 0.01$
for $m_{h_2}=20$~GeV and  $\sin^2\theta_s\simeq 0.1$ for
$m_{h_2}=100$~GeV~\cite{LEP:2003}.

\begin{figure}
  \begin{center}
    \includegraphics[width=0.7\linewidth]{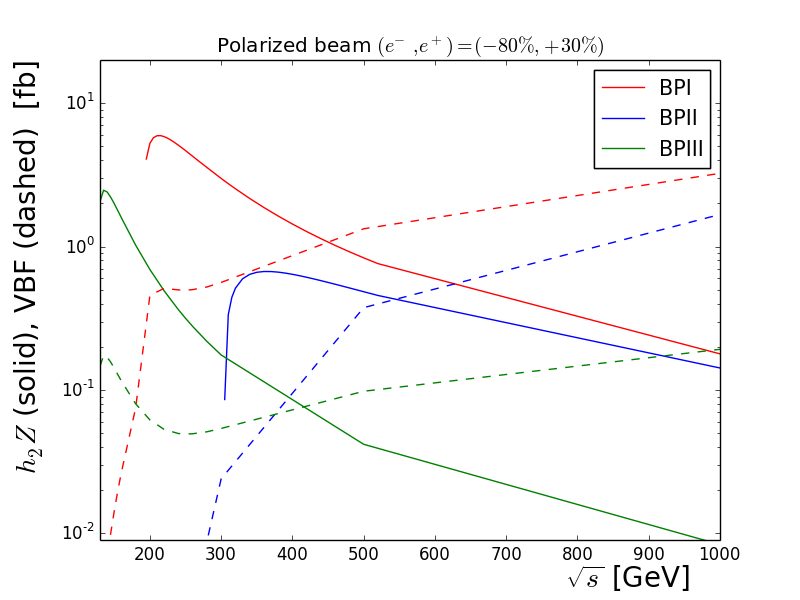}
  \end{center}
  \caption{
    The production cross sections for $h_2$ at various $\sqrt{s}$ in GeV with
    polarized beams.
    The solid curves correspond to the Higgsstrahlung process $e^+ e^-
    \to Z h_2$, and the dashed curves are the $WW$-fusion process
    $e^+ e^- \to \nu \bar{\nu} h_2$.
    The red, blue, and green colors denote the BP I, II, and III,
    respectively.\label{fig:eexsecm}
  }
\end{figure}

The bounds from the ILC experiment can significantly supersede the LEP
bounds due to the higher luminosity by a factor of a thousand as well
as polarized beams.
The main production processes for light Higgs bosons at the ILC are
Higgsstrahlung ($e^+ e^- \rightarrow Z h_2$) for small center-of-mass
energies and $WW$-fusion ($e^+ e^- \rightarrow \nu\bar{\nu} h_2)$ for
large center-of-mass energies.
By using \texttt{MadGraph 5}~\cite{Alwall:2014hca}, we obtain
the production cross sections of $e^+ e^- \to Z h_2$ and $e^+ e^- \to
\nu \bar{\nu} h_2$ for polarized beam $P(e^-, \, e^+) = (-80\%,
\, +30\%)$.
Fig~\ref{fig:eexsecm} shows the cross sections at various
center-of-mass energies.

Here we concentrate on the ILC run at $\sqrt{s}=250$~GeV
and apply the nominal results in Ref.~\cite{Drechsel:2018} for the
250~GeV with the integrated luminosity 2000~fb$^{-1}$, which are
extrapolated from the LEP results with the beam polarization $P(e^-,\,
e^+) = (-80\%, \, +30\%)$.
Two approaches, `ILC traditional' and `ILC recoil', are studied for
estimating the sensitivity of the ILC to light Higgs masses.
In the ILC traditional method, the signal process is identified with
light Higgs decay $h_2 \to b \bar{b}$ and $Z$ boson decay $Z \to \mu^+
\mu^-$.
On the other hand, in the ILC recoil method, only $Z$ boson decay ($Z
\to \mu^+ \mu^-$) is exploited and the recoil mass distribution is
studied.
The 95\% C.L. upper limits on the Higgs mixing angle
$\sin^2\theta_s$ are obtained for small Higgs masses below 125~GeV as
follows~\cite{Drechsel:2018}:
\begin{align}
  & \sin^2\theta_s \in [0.001, 0.002] \quad \rm{with \ ILC \
    traditional,} \label{eq:ILCtraditional} \\
  & \sin^2\theta_s \in [0.003, 0.005] \quad \rm{with \ ILC \
    recoil.\footnotemark}
\end{align}
\footnotetext{See Ref.~\cite{Wang:2018} for a more conservative estimation.}
For $m_{h_2} = 99.4$~GeV, the corresponding maximal reach at the ILC
is $\sin^2\theta_s \simeq 0.0015$.
Therefore, the signal of the light Higgs $h_2$ can be well
discovered in the case of the BP I, where $\sin^2\theta_s = 0.014$.

\subsection{BP II  (\boldmath$\psi \bar{\psi} \rightarrow h_1 h_1$ annihilation channel)}

The channel of DM annihilation into the SM Higgs pair ($h_1 h_1$ in
the SFDM) is an alternative possibility explaining the gamma-ray
excess. For example, a model independent study in
Ref.~\cite{Calore:2014nla} also showed that the DM annihilation into
the Higgs pair gives a good fit to the gamma-ray excess data if
$m_\text{DM} \simeq m_{h_1} \simeq 125.7$~GeV and the DM annihilation cross
section $\langle \sigma v \rangle \simeq 5.33\times
10^{-26}$~$\mathrm{cm}^3\,\mathrm{s}^{-1}$ for a self-conjugate DM\@.

In the SFDM model, this channel gives a best fit when $m_\psi =
127.5$~GeV,  $m_{h_1} = 124.9$~GeV, $m_{h_2} = 213.5$~GeV, and
$\sin\theta_s = - 0.11$ from the Lagrangian parameters $\lambda_0 =
0.1315$, $\lambda_1 =1237.8$~GeV, $\lambda_2 = - 2.0$, $\lambda_3 = -
820.5$~GeV, $\lambda_4 = 9.39$, $v_s = 306.15$~GeV, $g_S = 0.098$,
and $\sin\xi =1$~\cite{Kim:2016csm}.
From theses parameters, we obtain the correct DM relic density $\Omega
h^2 = 0.12$, and DM annihilation cross section $\langle \sigma v
\rangle = 2.11 \times 10^{-26}$~$\mathrm{cm}^3\,\mathrm{s}^{-1}$,
explaining the gamma-ray excess within the uncertainty of the galaxy
halo profile near the GC\@.
Note that the BP II is safe from various astrophysical bounds even when
the light Higgs $h_2$ is purely pseudoscalar in the dark sector. In the
following subsections, we discuss the collider signatures of this
benchmark point.

\subsubsection{Higgs signal strength}

In the BP II, the Higgs mixing angle is $\sin\theta_s=-0.11$, {\em i.e.},
$\cos\theta_s = 0.994$, which implies a $0.6\%$ deviation of $hZZ$
coupling from the SM value.
As commented in the previous subsection, the deviation is within the
reach of the 250~GeV ILC ($0.38\%$).

\subsubsection{Triple Higgs coupling}

The triple Higgs coupling $c_{111}=148.96$~GeV of the BP II
corresponds to 22\% deviation from the SM value.
The deviation can be marginally detected at the ILC with higher
center-of-mass energy whose accuracy is expected to be $\sim
20\%$~\cite{DiVita:2017vrr}.
On the other hand, it can be well measured in the 100~TeV $pp$
collider where $\sim 4\%$ accuracy might be reachable~\cite{CMS:2013xfa, ATLAS:HL-LHC-Higgs}.

\subsubsection{Exotic Higgs decays}

Since $m_{h_2} > m_{h_1}$ and $2m_\psi > m_{h_1}$, exotic Higgs decays
such as $h_1 \to h_2 h_2$ and $h_1 \to \psi \bar \psi$ are not
expected for the BP II\@.

\subsubsection{Production of {\boldmath$h_2$} at colliders}

The singlet-like Higgs mass of the BP II is $m_{h_2}= 213.5$~GeV.
For the mass region of $m_{h_1} < m_{h_2} < 2m_{h_1}$, $h_2$
dominantly decays to the pairs of $WW$ and $ZZ$. The
corresponding branching ratios are BR$(h_2 \to WW) = 71.5\%$ and
BR$(h_2 \to ZZ) = 28.0\%$ and the total decay width of the $h_2$ is
about 25~MeV.
The strongest constraint for the additional Higgs in this mass range comes from
the current LHC searches for a new scalar resonance decaying to a pair
of $Z$ bosons~\cite{CERN-PH-EP-2015-074, CERN-EP-2017-251,
  CERN-EP-2018-009}.
The 95\% C.L. upper bound on $\sigma(pp \to X \to ZZ)$ at the LHC with
an integrated luminosity of $\sim 36$~fb$^{-1}$ at the center-of-mass
energy of 13 TeV, is about $2.8 \times 10^{-1}$~pb for $m_X \simeq
210$~GeV.
That can be translated to the upper bound on the Higgs mixing angle
$\sin^2\theta_s\lesssim 0.06$ for $m_H \simeq 210$~GeV.
At the HL-LHC with integrated luminosity of 3000~fb$^{-1}$, an order
of magnitude improvement on the bound, $\sin^2\theta_s \lesssim 0.008$
at 95\% C.L., is expected~\cite{CMS-PAS-FTR-13-024}.
Therefore, we expect that the corresponding signal of $h_2$ with the mixing angle
$\sin^2\theta_s = 0.012$ can be observed at the HL-LHC\@.

\subsection{BP III (\boldmath$\psi \bar{\psi} \rightarrow h_2 h_2$ annihilation channel)}

Another viable scenario for the Fermi gamma-ray excess is the DM
annihilation process into dark sector particles.
The scenario is the most promising channel
explaining the gamma-ray excess in the sense that it is rather easy to
avoid various bounds from colliders and astrophysical observations.
The model independent study in Ref.~\cite{Dutta:2015} shows that the
DM annihilation into a pair of new scalars ($\phi\phi$), each of which
decays to a $b$-quark pair, can provide a good fit to the Fermi
gamma-ray excess data.
The best-fit is obtained for $m_\text{DM} \simeq 65$~GeV, the new
scalar mass is about the half of the DM mass $m_\phi = m_\text{DM}/2$,
and DM annihilation cross section $\langle \sigma v \rangle \simeq
2.45\times 10^{-26}$~$\mathrm{cm}^3\,\mathrm{s}^{-1}$, assuming a
self-conjugate DM\@.

In Ref.~\cite{Kim:2016csm}, it was shown that the above scenario can
be realized in the SFDM model framework with the model parameters as
follows.
In the SFDM, a best fit to the excess is obtained when $m_\psi =
69.2$~GeV, $m_{h_1} = 125.1$~GeV, $m_{h_2} = 35.7$~GeV and
$\sin\theta_s = 0.025$ from the parameters in the Lagrangian
$\lambda_0 = 0.13$, $\lambda_1 =4.5$~GeV, $\lambda_2 = - 0.0055$,
$\lambda_3 = - 391.51$~GeV, $\lambda_4 = 2.20$, $v_s = 276.21$~GeV,
and $g_S = 0.056$.
With these parameter setup, we obtain the
correct DM relic density $\Omega h^2 = 0.121$, and DM annihilation
cross section $\langle \sigma v \rangle = 2.26 \times
10^{-26}$~$\mathrm{cm}^3\,\mathrm{s}^{-1}$.
The benchmark scenario can explain the Fermi gamma-ray excess within
the uncertainty of the galaxy halo profile near the GC like the
other benchmark points.
We discuss the collider signatures of the BP III in this subsection.

\subsubsection{Higgs signal strength}

The Higgs mixing angle $\sin\theta_s=0.025$ corresponds to
$\cos\theta_s=0.9997$. It implies $0.03\%$ reduction of $hZZ$ coupling
from the SM value. The future colliders including the ILC are not
sensitive enough to measure such a small deviation.

\subsubsection{Triple Higgs coupling}

The triple Higgs coupling of the BP III is $c_{111} = 190.57$~GeV,
which implies $-0.15$\% deviation from the SM prediction.
Such a small deviation will be very difficult to measure even in the
100 TeV $pp$ collider.

\subsubsection{Exotic Higgs decays}

The singlet-like Higgs mass of the BP III ($m_{h_2}=35.7$~GeV) is
smaller than half of the SM-like Higgs mass, thus $h_1$ can decay into
a new scalar pair $h_2 h_2$.
The corresponding branching ratio is $\text{BR}(h_1 \to h_2 h_2) = 7\%$.
The light Higgs subsequently decays to mostly $b\bar{b}$,
with BR$(h_2 \to b\bar{b})=86\%$, giving rise to $\text{BR}(h_1 \to
h_2 h_2 \to 4b) = 5.2\%$.

Recently, the ATLAS collaboration provided a result of search for the
Higgs boson produced in association with a vector boson and decaying
into two spin-zero particles in the $H \to aa \to 4b$ channel at the
13~TeV LHC with an integrated luminosity of
36.1~fb$^{-1}$~\cite{ATLAS:CERN-EP-2018-128}.
The 95\% C.L. upper limit on the combination of cross sections for $WH$ and $ZH$ times the branching ratio of $H \to aa \to 4b$ ranges from 3.0~pb
for $m_a = 20$~GeV to 1.3~pb for $m_a = 60$~GeV. The upper limit is about 1.1~pb for $m_a=35.7$~GeV.
We can translate the result to an upper limit on the branching ratio $\text{BR}(h_1 \to h_2 h_2 \to 4b) \lesssim 50\%$ with $m_{h_2} = 35.7$~GeV
for the BP III, assuming the SM cross sections for $WH$ and $ZH$.
An order of improvement on the bound will be possible at the HL-LHC~\cite{Curtin:2013}.

Searches for exotic Higgs decays with final states involving quarks
are somewhat challenging at the LHC and HL-LHC\@.
However, the ILC at $\sqrt{s} = 250$~GeV has an excellent sensitivity
to search for such final states, due to low QCD backgrounds and the
recoil mass technique for tagging the Higgs boson associated with the
$Z$ boson.
Through the Higgsstrahlung process $e^+ e^- \to Z h_1$ and
the $h_1\to h_2 h_2 \to 4b$ decay,
the 250~GeV ILC with the integrated luminosity of 2~ab$^{-1}$ can
exclude branching ratio of $h_1 \to h_2 h_2 \to 4b$
down to $\sim 10^{-3}$~\cite{Liu:2016}.
Therefore, we expect that the exotic Higgs decay of the BP III will be
very well detected at the ILC experiment.

\subsubsection{Production of $h_2$ at colliders}

In the BP III, the Higgs mixing angle is $\sin^2\theta_s = 6.25 \times
10^{-4}$ and the mass of the singlet-like Higgs is $m_{h_2} = 35.7$~GeV.
Since the mixing angle value is much smaller than the ILC sensitivity given in
Eq.~(\ref{eq:ILCtraditional}), the signal for the $e^+e^-\to Zh_2$
process is beyond the reach of the ILC experiment.

\section{Conclusions\label{sec:concl}}

In this paper, we investigated various collider signatures of the three
benchmark points in the SFDM model.
The benchmark points were chosen to explain the Fermi gamma-ray excess
at the GC from the DM pair annihilations in the previous study.
According to the final state of the DM annihilation process, three
benchmark points have been considered: DM annihilations into a pair of
$b$ quarks, SM-like Higgs bosons, and new scalars, dubbed as BP I, BP
II, and BP III, respectively.
The probe of such signals at colliders is necessary in order to either
support or oppose the possibilities of explaining the gamma-ray excess
in terms of the DM annihilation and the SFDM model is a proper
reference model for that.
We need a unique strategy to explore the signals of the each benchmark point, representing a parameter set of the DM
annihilation channel providing a best fit to the observed excess.
We categorize the search strategies by listing the four observables:
Higgs signal strength, triple Higgs coupling, exotic Higgs decay, and
production of the new scalar, $h_2$, at future colliders such as ILC
(easily convertible to CEPC as well), HL-LHC, and FCC-hh.

In Table~\ref{table:summary}, we briefly summarize the preferred
observables to probe each benchmark point.
It turns out that the BP I is expected to be explored by the
measurements of the Higgs signal strength, exotic Higgs decay, and
$h_2$ production at the ILC\@.
As the triple Higgs coupling of the BP I is on the sensitivity limit
of the FCC-hh, we expect that observing the deviation of the triple
Higgs coupling will be marginally possible.
All the collider observables except the exotic Higgs decay can probe
the BP II\@. In particular, the triple Higgs coupling of the BP II is
substantially smaller than the SM prediction.
On the other hand, the BP III can be tested by the search for the exotic Higgs decay $h_1
\to h_2 h_2$. The direct production of $h_2$ is hardly measurable due
to the small mixing angle.
Since single collider observable is not sufficient to probe all the
benchmark scenarios simultaneously, combined searches must be
performed to find the new-physics signals at future lepton and hadron
colliders.

\begin{table}[t!]
  \centering
  \scriptsize
  \begin{tabular}{>{\centering}m{0.9cm}||>{\centering}m{2.8cm}|>{\centering}m{2.8cm}|>{\centering}m{3.1cm}|>{\centering}m{3.6cm}}
    \hline
    & Higgs signal strength
    & Triple Higgs coupling
    & Exotic Higgs decay
    & $h_2$ production
    \tabularnewline
    \hline
    BP I
    & 0.7\% reduction \newline $\bigcirc$ \newline ($\delta \sim
      0.3$\% (ILC))
    & 3.9\% reduction \newline $\bigtriangleup$ \newline ($\delta \sim
      4$\% (FCC-hh))
    & BR$(h_1 \to \psi \bar\psi) = 0.73$\% \newline $\bigcirc$
      \newline (BR$(h \to \text{invisible}) \lesssim 0.3$\% (ILC))
    & $m_{h_2} = 99.4$~GeV, \newline $\sin^2 \theta_s = 0.014$ \newline
      $\bigcirc$ \newline ($\sin^2 \theta_s \lesssim 0.0015$ (ILC))
    \tabularnewline
      \hline
    BP II
    & 0.6\% reduction \newline $\bigcirc$ \newline ($\delta \sim
      0.3$\% (ILC))
    & 22\% reduction \newline $\bigcirc$ \newline ($\delta \sim
      4$\% (FCC-hh))
    & No exotic decay \newline \xmark
    & $m_{h_2} = 213.5$~GeV, \newline $\sin^2 \theta_s = 0.012$ \newline
      $\bigcirc$ \newline ($\sin^2 \theta_s \lesssim 0.008$ (HL-LHC))
    \tabularnewline
      \hline
    BP III
    & 0.03\% reduction \newline \xmark \newline ($\delta \sim
      0.3$\% (ILC))
    & 0.15\% reduction \newline \xmark \newline ($\delta \sim
      4$\% (FCC-hh))
    & BR$(h_1 \to h_2 h_2) = 7$\% \newline $\bigcirc$ \newline
      (BR$(h \to s s \to 4b) \lesssim 0.1$\% (ILC))
    & $m_{h_2} = 35.7$~GeV, \newline $\sin^2 \theta_s = 6.3\times
      10^{-4}$ \newline \xmark \newline ($\sin^2 \theta_s \lesssim
      0.0015$ (ILC))
    \tabularnewline
    \hline
  \end{tabular}
  \caption{Summary of the collider signatures for the benchmark points.
    The circle (triangle) denotes that the collider signal of the
    benchmark can be (marginally) measurable at future colliders.
    The cases where the collider signal is expected to be hard to
    measure or beyond the reach of the future colliders are marked
    with the cross.
    The texts in parentheses are relevant bounds expected at the
    future colliders.\label{table:summary}}
\end{table}

\section*{Acknowledgments}
We would like to thank Jia~Liu, Zhen~Liu, and Carlos~Wagner for useful
discussions.
This work was performed in part at the Aspen Center for Physics, which
is supported by National Science Foundation grant PHY-1607611.
YGK is supported by the Basic Science Research Program through the
National Research Foundation of Korea (NRF) funded by the Korean
Ministry of Education, Science and Technology
(NRF-2018R1D1A1B07050701).
The work of CBP is supported by IBS under the project code,
IBS-R018-D1.
SS is supported by the National Research Foundation of Korea
(NRF-2017R1D1A1B03032076).
SS appreciates the hospitality of Fermi National Accelerator
Laboratory.

\end{document}